\documentclass[twocolumn,prl,aps,superscriptaddress,showpacs,
nofootinbib,preprintnumbers,floats,floatfix]{revtex4}

\usepackage{amsmath}
\usepackage{amssymb}
\usepackage{latexsym}
\usepackage{graphicx}
\usepackage{hyperref}
\usepackage{bm}

\newcommand{\fnl}{\ensuremath{f_{\mathrm{NL}}}}
\newcommand{\Nf}{N_{\mathrm{f}}}
\newcommand{\Nh}{\bar{N}}
\newcommand{\Mp}{M_{\mathrm{P}}}
\newcommand{\Ps}{\mathcal{P}}

\newcommand{\alternative}[2]{#1}

\newcommand{\arxiv}[1]{\href{http://arxiv.org/abs/#1}{#1}}
\newcommand{\arxivnew}[1]{\href{http://arxiv.org/abs/#1}{arXiv:#1}}

\begin{document}

\title{Non-gaussianity in axion N-flation models}

\author{Soo A.~Kim} 
\affiliation{Department of Astronomy and Space Science, Kyung Hee
  University, Yong-in 446-701, South Korea}   
\author{Andrew R.~Liddle} 
\affiliation{Astronomy Centre, University of Sussex,
  Brighton BN1 9QH, United Kingdom} 
\author{David Seery}
\affiliation{Astronomy Centre, University of Sussex, Brighton BN1 9QH,
  United Kingdom} 
\date{\today}

\begin{abstract}
We study perturbations in the multi-field axion N-flation model, taking
account of the full cosine potential.  We find significant differences
with previous analyses which made a quadratic approximation to the
potential. The tensor-to-scalar ratio and the scalar spectral index move
to lower values, which nevertheless provide an acceptable fit to
observation.  Most significantly, we find that the bispectrum
non-gaussianity parameter $\fnl$ may be large, typically of order 10
for moderate values of the axion decay constant, increasing to of
order 100 for decay constants slightly smaller than the Planck scale.
Such a non-gaussian fraction is detectable. We argue that this
property is generic in multi-field models of hilltop inflation.
\end{abstract}


\maketitle


Much focus has been placed lately on the discovery potential of cosmic
non-gaussianity in the statistics of primordial perturbations. The
Wilkinson Microwave Anisotropy Probe (WMAP) has already set
interesting limits \cite{WMAP7}. The Planck satellite, now taking
data, will improve these significantly, reaching a sensitivity to the
non-gaussianity parameter $\fnl$ of around five. Discovery of
non-gaussianity would open a new arena of cosmological observations
particularly suited to probing early Universe physics.

Present ideas in fundamental physics suggest there may be many scalar
fields which can influence the early Universe, including
inflation. N-flation \cite{DKMW} uses many string axions to provide a
realization of the `assisted inflation' phenomenon \cite{LMS}, in
which a collection of scalar fields cooperatively support inflation
even if their potentials are individually too steep.  The
phenomenology of such models (see also Ref.~\cite{mfield}) links
fundamental physics and upcoming cosmological observations.

Previous N-flation studies have assumed that all relevant fields are
close to their minima and can be described by quadratic
potentials. For axions the full potential is trigonometric and we find
the quadratic approximation is unreliable. Even for identical
potentials, the condition for stable co-evolution of the fields is
violated near the hilltop \cite{CL}.  Therefore fields in this region
evolve on divergent trajectories.  Accounting for this divergence by
retaining the full potential leads to two very significant
changes. The predicted scalar spectral index and tensor-to-scalar
ratio, $r$, are reduced. This remains compatible with existing
observations but may leave $r$ undetectable. More importantly, $\fnl$
is predicted to be large, and very plausibly within the range of
future probes.

This unexpectedly large non-gaussianity is a genuine multi-field
phenomenon. It is a consequence of the diverging trajectories near the
hilltop, implied by a negative $\eta$-parameter of order unity or
larger.  In single-field models, potentials of this form lead to a
density perturbation with a spectral index, $n$, in conflict with
observation. The assisted inflation mechanism reduces $1-n$ to an
acceptable value, but leaves $\fnl$ dominated by the
contribution of the field closest to the peak.

\section{The model}

The axion N-flation model is based on a set of $\Nf$ uncoupled fields,
labelled $\phi_i$, each with a potential \cite{DKMW}
\begin{equation}
	V_i 
	= \Lambda_i^4 \left ( 1- \cos \alpha_i \right) \,,
\end{equation}
where $\alpha_i = 2\pi\phi_i/f_i$ and $f_i$ is the $i^{\mathrm{th}}$
axion decay constant.  In a more general model couplings may exist
between the fields, but we will not consider these.  The mass of each
field in vacuum satisfies $m_i = 2\pi \Lambda_i^2/f_i$, and the
angular field variables $\alpha_i$ lie in the range
$(-\pi,+\pi]$. Without loss of generality we will set initial
conditions with all $\alpha_i$ positive.  If only a single field is
present this model is known as natural inflation \cite{natural}.

Calculation of the observables $n$, $r$ and $\fnl$ makes use of the
$\delta N$ formula \cite{deltaN}, which considers how the total number
of $e$-folds of expansion $N$ is modified by field perturbations.  We
define slow-roll parameters for each field as
\begin{equation}
	\epsilon_i \equiv \frac{\Mp^2}{2}
	\left( \frac{V_i'}{V_i} \right)^2\,,
	\label{eq:epsi}
\end{equation}
where $\Mp \equiv (8 \pi G)^{-1/2}$ is the reduced Planck mass, a
prime denotes the derivative of a function with respect to its
argument, and no summation over $i$ is implied.  The global slow-roll
parameter $\epsilon \equiv -\dot{H}/H^2$ can be written as a weighted
sum $\epsilon \simeq \sum_i (V_i/V)^2 \epsilon_i$, in which each field
contributes according to its share of the total energy density. We
must have $\epsilon < 1$ during inflation.

We work in the horizon-crossing approximation, in which the dominant
contribution to each observable is assumed to arise from fluctuations
present only a few $e$-folds after horizon exit of the wavenumber
under discussion.  After smoothing the universe on a superhorizon
scale somewhat smaller than any scale of interest, the
horizon-crossing approximation becomes valid whenever the ensemble of
trajectories followed by smoothed patches of the universe approaches
an attractor.  We suppose that inflation exits gracefully, with each
field settling into the minimum of its potential.  The
horizon-crossing formulas will then be a reasonable approximation.
Using this method, and conventional definitions for each observable
parameter \cite{PDP}, we find
\begin{eqnarray}
\Ps_\zeta & = & \frac{H_*^2}{4\pi^2} \sum_i N_{,i}N_{,i} =
\frac{H_*^2}{8\pi^2 \Mp^2} \sum_i \frac{1}{\epsilon_i^*}\,;
\label{eq:P} \\
n-1 
& = &-2\epsilon_* - \frac{8\pi^2}{3H_*^2} \sum_j \frac{\Lambda_j^4}{f_j^2} 
    \frac{1}{\epsilon_j^*} \Big/ \sum_i
    \frac{1}{\epsilon_i^*} \,; \label{eq:n} \\  
r & = & \frac{2}{\pi^2 \Ps_\zeta} \frac{H_*^2}{\Mp^2} = 16 \Big/ \sum_i
\frac{1}{\epsilon_i^*}\,; \label{eq:r} \\
\frac{6}{5} \fnl & \simeq &
\frac{\sum_{ij}N_{,i}N_{,j}N_{,ij}}{\left(\sum_k N_{,k}N_{,k}\right)^2}
= \frac{r^2}{128} \sum_i \frac{1}{\epsilon_i^*} \frac{1}{1+\cos
  \alpha_i^*} \,,
  \label{eq:fnl}
\end{eqnarray}
where $N_{,i}$ and $N_{,ij}$ are respectively the first and second
derivatives of $N$ with respect to the fields, and $\ast$ indicates
evaluation at horizon crossing (determined by Eq.~(\ref{eq:ntot})
below).  In writing Eq.~\eqref{eq:fnl} any intrinsic non-gaussianity
among the field perturbations at horizon crossing has been neglected,
a good approximation provided $\fnl > 1$ \cite{Maldacena,VW}.  Our
sign convention for $\fnl$ matches WMAP \cite{WMAP7}, and the
non-gaussianity is predicted to be of local type. The observed
amplitude of perturbations is obtained by adjusting the $\Lambda_i$ to
give an appropriate value of $H_*$.

Under a quadratic approximation to each potential, it can be shown
that Eqs.~\eqref{eq:r} and \eqref{eq:fnl} recover their single-field
values of order $\sim 1/N_\ast$ \cite{single,VW}, making $\fnl$
undetectably small.  The spectral index can be shown to be less than
its single field value $1-2/N_*$ \cite{LR} with equality only in the
equal-mass case. Its value for a given choice of parameters must be
computed numerically \cite{KLs}.  However, we will see that these
results all change whenever our initial conditions populate the
hilltop region.

\section{N-flation perturbations}

Eqs.~\eqref{eq:P}--\eqref{eq:fnl} apply for any choice of $\Lambda_i$
and $f_i$. We restrict attention to the case where all fields have the
same potential, which already captures the interesting
phenomenology. A broader investigation will be published
elsewhere. The scale $\Lambda \equiv \Lambda_i$ is fixed from the
observed amplitude of $\Ps_\zeta$, leaving $f \equiv f_i$ and $N_{\rm
  f}$ as adjustable parameters.  The initial conditions are drawn
randomly from a uniform distribution of angles $\alpha_i$, with
several realizations to explore the probabilistic spread. From these
two parameters we predict the observables $n$, $r$ and $f_{\rm NL}$.

There are two constraints. First, we require sufficient
$e$-foldings. For a given set of initial angles $\alpha_i$, and
ignoring a small correction from the location of the end of inflation,
one finds
\begin{equation}
N_{\mathrm{tot}} \simeq
\sum_i
\left(\frac{f_i}{2\pi \Mp}\right)^2 \ln \frac{2}{1+\cos \alpha_i}
\simeq
\frac{\ln 2}{2\pi^2} \frac{f^2}{\Mp^2} \Nf \,,
\label{eq:ntot}
\end{equation}
where in the second equality we have replaced $N_{\mathrm{tot}}$ with
its expectation value by averaging over
$\alpha_i$. Eq.~\eqref{eq:ntot} is replicated to high accuracy in
numerical simulations. For a given $f$ it determines the minimum
number of fields required for sufficient inflation, typically
several hundred or more.  There is no similar constraint from the
spectral index.  When $N_{\mathrm{tot}} \approx N_\ast$, the
$\alpha_i^\ast$ are uniformly distributed and $\langle n-1 \rangle
\simeq -5 \ln 2 / N_{\ast}$, independent of $f$ and $N_{\rm f}$.  This
tilt is observationally acceptable.  For larger $\Nf$ the spectral
index approximately satisfies Eq.~\eqref{eq:napprox} below.

Second, a key motivation of the N-flation model was to obviate the
requirement for superplanckian field values, which are invoked in many
single-field models. If one literally imposes $|\phi| < \Mp$ this
requires $f_i<2\Mp$ for each $i$. However, it would be reasonable to
regard this condition as approximate and not mandatory.  

The $\epsilon_i$ approach zero for fields close to the hilltop, so
each summation in Eqs.~\eqref{eq:P}--\eqref{eq:fnl} is dominated by
those fields with the smallest $\epsilon_i$.  Suppose some number
$\Nh$ of fields have roughly comparable $\epsilon_i$, of order
$\bar{\epsilon}$. The observable parameters have different scalings
with $\Nh$. The spectrum, $\Ps_\zeta$, scales like $\Nh$ copies of a
single-field model with slow-roll parameter $\bar{\epsilon}$, whereas
$r$ is reduced by a factor $\Nh$ compared to its value in the same
single-field model. The spectral index can be written exactly (within
slow-roll) in terms of a single sum coming from $H_*$,
\begin{equation}
	n - 1 \approx - 2 \epsilon_\ast - 8 \pi^2
		\left( \frac{\Mp}{f} \right)^2 \Big/
		\sum_i (1 - \cos \alpha_i^\ast) \,,
	\label{eq:napprox}
\end{equation}
and is independent of $\Nh$. It becomes close to $-2\epsilon_\ast$
when the denominator is of order $10^3$. This is the standard
assisted-inflation mechanism.  Most importantly, $\fnl$ has the
approximate behaviour
\begin{equation}
	\frac{6}{5} \fnl \approx \frac{2\pi^2}{\Nh}
	\left( \frac{\Mp}{f} \right)^2 ,
	\label{eq:fnlapprox}
\end{equation}
which is independent of $\bar{\epsilon}$ if the dominant fields are
sufficiently close to the hilltop.  N-flation has lifted the
single-field consistency condition $\fnl \approx - (5/12)(n-1)$
\cite{Maldacena,VW}, which prevents single-field models generating
large non-gaussianity without violating observational bounds on $n$.

Where the summations in Eqs.~\eqref{eq:P}--\eqref{eq:fnl} are
dominated by a single field, this formula shows that $\fnl$ can become
rather large, scaling as $(\Mp/f)^2$. For $f = \Mp$, we find $\fnl
\lesssim 16.4$; a non-gaussian fraction of this magnitude should be
visible to the Planck satellite.  It is even possible to achieve $\fnl
\sim 100$ for $f \sim 0.4 \Mp$, though then $N_{\rm
  f}$ must be very large to gain sufficient $e$-foldings.  If $\fnl
\gtrsim 50$ it may be more profitable for Planck to search for
non-linearity in the trispectrum \cite{trispec}, for which estimates
in the quadratic approximation were given in Ref.~\cite{SL}. We defer
a full analysis of the trispectrum to future work but note that the
trispectrum equivalents of Eq.~(\ref{eq:fnlapprox}) are, in
conventional notation \cite{trispectrum}, $\tau_{\rm NL} =
(4\pi^4/\Nh^2) (\Mp^4/f^4)$ and $(54/25)g_{\rm NL} = (8\pi^4/\Nh^2)
(\Mp^4/f^4)$.

\begin{figure}[t]
\includegraphics[width=6 cm,angle=90]{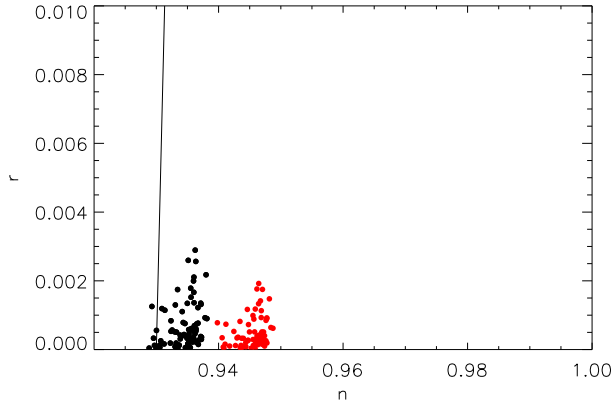}\\
\caption{Predictions in the $n$--$r$ plane, averaged over
  realizations, for various values of $f$ between $0.4\Mp$ and $2\Mp$
  and of $\Nf$ between 464 and 10,000, all giving sufficient
  inflation. The black (left) cluster of points takes $N_* = 50$ and
  the red (right) cluster $N_*=60$. The quadratic expansion predicts
  $r=8/N_*$, far off the top of this plot.  The region right of the
  line is within the WMAP7+BAO+$H_0$ 95\% confidence
  contour~\cite{WMAP7}.
  \label{f:nr}}
\end{figure}

\begin{figure}[t]
\includegraphics[width=6 cm,angle=90]{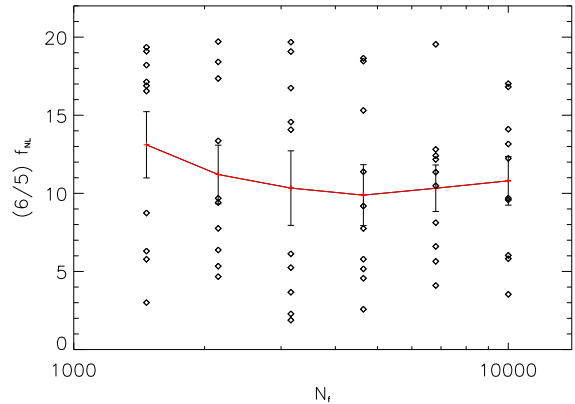}\\
\caption{Predicted non-gaussianity, $\frac{6}{5}\fnl$, for $f=\Mp$ and
  $N_*=50$. The error bars are on the mean over realizations (not the
  standard deviation). Here the maximum achievable value of
  $\frac{6}{5} \fnl$ is $2\pi^2 \simeq 20$, almost saturated in some
  realizations. The significant spread is due to initial condition
  randomness with typical mean values being around half the maximum
  achievable value, and no discernable trend with $N_{\rm
    f}$. \label{f:fnl}}
\end{figure}

\begin{figure}[t]
\includegraphics[width=6 cm,angle=90]{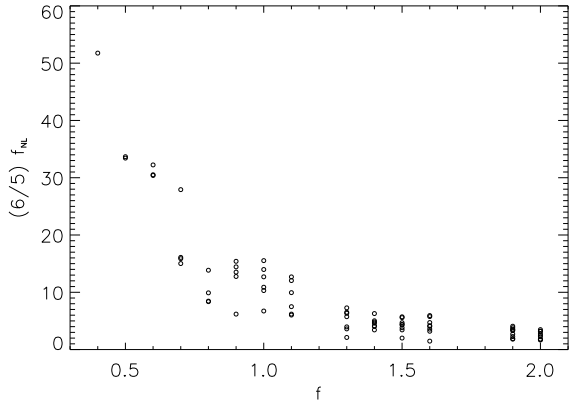}\\
\caption{The predicted non-gaussianity as a function of $f$, for a
  range of choices of $N_{\rm f}$. Each point shown is the average of
  five or more realizations for an $f$--$N_{\rm f}$ pair. We see a
  strong trend with $f$, well represented by Eq.~(\ref{eq:fnlapprox})
  with $\Nh \simeq 2$. The different $N_{\rm f}$ are scattered by
  randomness in the initial conditions rather than an identifiable
  trend.\label{f:fnl2}}
\end{figure}

The expectations described above are borne out in numerical
calculations.  In Fig.~\ref{f:nr} we show model predictions in the
$n$--$r$ plane, averaged over several realizations of the initial
conditions. We see $n$ and $r$ are only weakly dependent on the model
parameters (though there is significant dispersion amongst
realizations, not shown here), with the choice of $N_*$ being the
principal determinant of $n$.  In Fig.~\ref{f:fnl} we plot $\fnl$ as a
function of $\Nf$ for $f = \Mp$, with ten realizations at each $N_{\rm
  f}$. This clearly shows the expected maximum, which is nearly
saturated in cases where a single field dominates the summations. In
cases where several fields contribute significantly to the sums in
Eqs.~\eqref{eq:P}--\eqref{eq:fnl}, the non-gaussian fraction is
reduced. Fig.~\ref{f:fnl2} shows the mean predicted non-gaussianity,
averaged over realizations, as a function of $f$.

Eqs.~\eqref{eq:napprox} and~\eqref{eq:fnlapprox} clarify the origin of
large $\fnl$ in this model. The cooperative effect of the N-flation
mechanism does not enhance the non-gaussian signal. Indeed, $\fnl$ is
suppressed by the central limit theorem where $\Nh \gg 1$ fluctuations
contribute equally to the curvature perturbation. Nor does the large
effect arise from a singularity in the $e$-folding history, $N$, as a
function of its initial angles $\alpha_i$. Although
Eq.~\eqref{eq:ntot} is singular in the limit $\alpha_i \rightarrow
\pi$, its Taylor expansion is trustworthy unless $|\alpha_i - \pi|
\lesssim (\Ps_\zeta r)^{1/2}(\Mp /f_i)$. The observed magnitude of
$\Ps_\zeta$ requires $|\alpha_i - \pi| \gtrsim r^{1/2} (f_i/\Mp)$ for
each field, so a breakdown of the Taylor expansion cannot become
relevant unless at least one $f_i$ is a few orders of magnitude less
than the Planck scale, of order $(f_i / \Mp)^4 \lesssim \Ps_\zeta$.
These constraints additionally imply that we do not trespass on any
region of field space where quantum diffusion competes with classical
motion.

Instead, the large $\fnl$ derives from a generic dispersive effect
present in any hilltop potential.  Measuring the displacement of
$\phi_i$ from the hilltop by $\delta_i$, each potential can be
approximated in its vicinity by $V_i \approx 2 \Lambda^4_i
( 1 + \eta_i \delta_i^2 / 2\Mp^2 )$, where $\eta_i < 0$ satisfies
\begin{equation}
	\eta_i \equiv \Mp^2 \frac{V_i''}{V_i} \simeq
	- 2 \pi^2 \left( \frac{\Mp}{f_i} \right)^2 .
\end{equation}
These potentials are tachyonic.  Fields close to the hilltop remain
almost stationary, while fields further away are ejected downhill.
This process typically leaves a few fields on top of the hill, which
have small $\epsilon_i$ and dominate the sums in
Eqs.~\eqref{eq:P}--\eqref{eq:fnl}.  It seems clear this behaviour is
generic for any N-flation model constructed using hilltop potentials.
The few fields remaining in the vicinity of the hilltop each generate
contributions to the curvature perturbation with third moment $(6/5)
\fnl \approx - \eta_\ast$ \cite{Maldacena}.  Accounting for
suppression arising from the central limit theorem, we recover the
approximate expression~\eqref{eq:fnlapprox}.  For a general hilltop
potential, well-rehearsed arguments lead us to expect $|\eta| \sim 1$
and therefore $\fnl \sim 1$.  In a single-field model this is the
`$\eta$ problem'. In an N-flation model, it is a generic expectation
of enhanced non-gaussianity.  Even larger yields are possible in some
models, including our case, if it is possible to achieve $|\eta| \gg
1$ while preserving technical naturalness.

\section{Conclusions}

We have described a new mechanism for generating observably-large
cosmic non-gaussianity, based on the strongly-dispersive dynamics of
fields in a hilltop region. In a multi-field context such as the axion
N-flation model, assisted inflation can yield a viable spectral index
without a major dilution of the non-gaussianity. As compared to the
quadratic potential approximation to N-flation, we found a substantial
decrease in $r$, a modest increase in $1-n$, and a substantial
increase in $\fnl$.  These changes will happen whenever initial
conditions have a significant probability of populating the hilltop
region, such as the uniform (in field angle) initial conditions we
chose.

Searches have previously been made for models which achieve $|\fnl|
\gg 1$ while preserving slow-roll during inflation \cite{BCH}. The
N-flation model is of this type, but offers several advantages.  The
non-gaussian fraction is naturally bounded above, so that $\fnl$
cannot become arbitrarily large.  Therefore our predictions do not
depend on a sudden exit from inflation, e.g.\ triggered by a hybrid
transition, to prevent $\fnl$ from growing to an unacceptable value.
Equally important, our large signal does not derive from a singularity
of the $e$-folding history $N$, as a function of its initial
conditions.  These means we can rely on a perturbative expansion.  
We can simultaneously satisfy observational constraints on the
spectral index and tensor fraction.  Moreover, this result seems
generic. Inflation is self-replicating on top of the hill, sometimes
described as `topological inflation' \cite{topo}.  Coupled with the
dispersion of trajectories originating from the vicinity of the
hilltop, this implies large non-gaussianity may not be uncommon over a
landscape of scalar field vacua.


\begin{acknowledgments}
S.A.K.\ was supported by NRF Grant No.\ 2009-0078118 and
A.R.L.\ and D.S.\ by the Science and Technology
Facilities Council [grant number ST/F002858/1].  S.A.K.\ thanks KASI,
APCTP and Sussex for hospitality, and A. Moraghan, M.-R. Kim, and
S. Kim for help with figures.
\end{acknowledgments}


\end{document}